\documentstyle[epsf,12pt]{article}

\begin{document}
\renewcommand{\baselinestretch}{1.5}

\newcommand\beq{\begin{equation}}
\newcommand\eeq{\end{equation}}
\newcommand\bea{\begin{eqnarray}}
\newcommand\eea{\end{eqnarray}}

\newcommand\al{\alpha}
\newcommand {\dlt}{ \frac{\delta}{\pi}}
\newcommand {\dlts}{\frac{\delta^2}{\pi^2}}

\newcommand\pisig{\Pi_{\sigma i }}
\newcommand\pisigm{\Pi_{I\sigma i }}
\newcommand\sumsig{\sum_{\sigma i }}
\newcommand\psig{p_{\sigma i }}
\newcommand\xsig{x_{\sigma i }}
\newcommand\xmsig{x_{-\sigma j }}
\newcommand\sumI{\sum_{I}^{N}}

\newcommand\sumi{\sum_i}
\newcommand\sumj{\sum_j}
\newcommand\sumJ{\sum_J}

\newcommand\expo{e^{iS(\{x_{+i}\}, \{x_{-i}\})}}
\newcommand\expon{e^{-iS(\{x_{+i}\}, \{x_{-i}\})}}

\newcommand\tp{\tilde\phi}
\newcommand\dpi{\delta/\pi}
\newcommand\dtpi{\delta/2\pi}
\newcommand\ddpi{\delta^2/\pi^2}
\newcommand\px{\partial_x}
\newcommand\prl{Phys. Rev. Lett.}
\newcommand\prb{Phys. Rev. {\bf B}}
\hfill MRI-PHY/P9805046 

\centerline{\bf Exactly Solvable Fermionic $N$-chain  Models}
\vskip 1 true cm

\centerline{Ranjan Kumar Ghosh}
\centerline{\it Haldia Government College, P.O. Debhog, }
\centerline{\it Midnapore 721657, India.}
\vskip .5 true cm

\centerline{P. K. Mohanty \footnote{{\it e-mail
address}: peekay@mri.ernet.in}and Sumathi Rao \footnote{{\it e-mail
address}: sumathi@mri.ernet.in}}  
\centerline{\it Mehta Research Institute, Chhatnag Road,Jhunsi,}
\centerline{\it Allahabad 211019, India.}

\vskip 2 true cm
\noindent {\bf Abstract}
\vskip 1 true cm

Motivated by the problem of $N$ coupled Hubbard chains, we investigate
a generalisation of the Schulz-Shastry model containing two species
of one-dimensional fermions interacting via a gauge field that depends
on the positions of all the particles of the other species. The exact 
many body ground state of the model can be easily obtained through a 
unitary transformation of the model. The correlation functions are
Luttinger-like - $i.e.$, they decay through power laws with
non-integer exponents. Through the interaction dependent
correlation functions of the two-particle operators, we identify the
relevant perturbations and hence, possible instabilities.
 
\vskip 1 true cm


\newpage

Exactly solvable models\cite{EXACT,LUTTINGER,LM} 
have always attracted a lot of interest
in theoretical physics, because they serve as paradigms for more
complicated systems. The fact that these models are usually in one
dimension no longer make them unrealistic, however, since current
technological advances have seen the advent of many semi-artificial
one dimensional systems, such as quantum wires, quantum Hall bars,
one-dimensional organic metals and one dimensional spin chains\cite{HA}.
In fact, phenomena such as one dimensional Luttinger liquids and the 
Haldane gap in spin chain models have actually been experimentally
seen\cite{AZUMA}. Besides their role in these systems, exactly solvable
models have played a very important role as  a reliable test
for various approximation methods and for developing qualitative
understanding\cite{EMERY}.

However, for two or more dimensions, there have been very few exact results.  
For instance, the large $U$ Hubbard model has been studied
using several approximation schemes\cite{2DHUB}, none of which  have
led to completely reliable results. 
In recent years, there have  been attempts to understand two dimensions
through the coupling of one-dimensional chains. Both coupled spin
chains\cite{SCHULZ} and coupled Hubbard models\cite{CHM} with
interchain hopping and interchain interactions
have been studied using a variety 
different schemes such as  
weak coupling renomalisation
group techniques  and bosonisation\cite{FISHER}, 
exact numerical diagonalisations\cite{NUM}, etc. 
Unfortunately,
in the absence of any exact results, the interpretation of the 
results of these
inter-chain coupling studies have remained difficult\cite{STRONG}.

With the motivation of approaching two dimensional phenomena through
the coupling of one-dimensional chains,
in this paper, we study a generalisation of a class of models\cite{SS}
that can
be diagonalised by a pseudo-unitary transformation and still exhibit
non-trivial Luttinger liquid behaviour. Our model has two species of
particles with pseudospin index $\sigma = \pm$, at the positions
$\xsig$ and with momenta $\psig$,  with a 
Hamiltonian given by
\beq
H =  \sumI \sumsig a_{I} (\pisig)^{2I}.
\label{one}
\eeq
Here, $\pisig = \psig + \sigma A_{\sigma}(\xsig)$ is the `covariant
momentum' introduced in Ref.\cite{SS} and  $N$ is a `chain
index' or `band index'. (The nomenclature of chain or band index will
be explained later.) 
We have chosen to have only even powers of the covariant momentum in the
Hamiltonian, although positivity of energy only requires that the
largest power of the covariant momentum be even. This maintains
the symmetry $x \rightarrow -x$ or parity, 
which simplifies the presentation of the calculations, although the
result goes through even when we include odd powers of the momenta. 
As explained in Ref.\cite{SS}, particles interact via a
gauge potential, given for the particle at the position $x$ by
$A_{\sigma}(x) =\sumj V(x-\xmsig)$ - $i.e.$, the potential for the
particles with positive pseudospin  is due to the presence of the particles
with negative pseudospin and vice-versa. The potential is chosen to be an
even function, vanishes at infinity and explicitly breaks time-reversal
invariance, although it is invariant under a combined operation of 
time reversal
and reversal of pseudo-spin index. We have generalised the model in
Ref.\cite{SS} by
including a chain index (or equivalently a band index) and allowing 
higher powers of the covariant momentum in the Hamiltonian.
Our model reduces to the Schulz-Shastry model for $I=1$ and $a_1 = 1$.
Clearly, the $a_{I}$ are not dimensionless, and in fact, explicitly
contain a scale $\Lambda$ (except for $a_1$, which is dimensionless).  
  
As noted by the authors in Ref.\cite{SS}, the same pseudo-unitary 
transformation that they use to diagonalise their Hamiltonian,
\beq
\expo p_{\sigma i }\expon = p_{\sigma i} - \partial_{x_{\sigma i}}
S(\{x_{+i}\}, \{x_{-i}\})
\eeq
diagonalises any power of $\pisig$, as long as we choose the function 
$S$ (a function of the $2n$ positions of the particles) to
eliminate the interaction in Eq.(\ref{one}).
Thus, we obtain the transformed Hamiltonian given by   
\beq
{\tilde H} = \expo H \expon =  \sumsig \sumI a_{I} 
(\psig)^{2I}=\sumsig H_{\sigma i}
\label{nham} 
\eeq
where the interaction pieces have been removed by the transformation.
However, the eigenvalues and eigenfunctions are not the same as that
for a genuinely non-interacting Hamiltonian because the boundary
conditions on the wave-functions are now different.

For the single particle Hamiltonian $H_{\sigma i}$ in Eq.(\ref{nham}),
the eigenvalue equation is a $2N^{th}$ order differential equation,
and depending on the energy chosen, will have at most $2N$ different
solutions. 
The general solution is given by
\beq
\tilde \psi = \sumI c_I e^{ik_I x} + h.c.,
\eeq
where the $k_I$'s are known in terms of E and the 
$N-1$ constants $a_I$ (we always
choose $a_N =1$ without loss of generality since it only sets the
overall scale).
However, not all $k_I$'s need be integer multiples of $2\pi/L$, where
$L$ is the size of the system. For
those that are not, the
corresponding $c_I$ vanish so as to make the wave function periodic in
$L$. Since the $a_I$'s are fixed, we may choose only one of the $k_I$'s to
be independent, say $k$,  which in turn fixes the dispersion to be
\beq 
E(k)=\sumI a_{I}(k^{2})^{2I}.
\label{disp}
\eeq
The Fermi points are the roots of the equation $E(k)=E_F$. We
choose an energy $E_F$ where the $2N$ roots $\{-b_I,b_I\}$ are all
real and distinct (with $b_1<b_2<\cdots <b_N$). The Fermi points
$\{-F_I, F_I\}$ are given by $F_{I} =2\pi [r_{I}]/L$ where $b_I=2\pi
r_I/L$ and $[r_i]$ stands for the largest integer below $r_I$.

We now see the justification for calling $I$ the chain or band index. The 
usual identification of an $N$-chain model with an $N$-band model is
made by
diagonalising the kinetic energy of the $N$-chain model and using the
$N$ different momenta along the direction perpendicular to the chain
to label the $N$ bands or $N$ dispersion relations\cite{FISHER}. The
filling of these bands upto the Fermi level defines the set of $2N$  
Fermi points $\pm k_F^I$. Our model is slightly different from the
$N$-chain model in that it  has only a single
dispersion(Eq.(\ref{disp})). However, the dispersion is not quadratic
and has $N$ wells (unlike the usual quadratic dispersion which has one
well per band) and $2N$ Fermi points. Hence, the physics it describes
is similar to that of the $N$ band model.

We are
interested in the solution of the original Hamiltonian in
Eq.(\ref{one}) 
and not the transformed Hamiltonian in Eq.(\ref{nham}).
Although the single particle energies of the two
Hamiltonians are the same, their  wave-functions 
are related by the pseudo-unitary transformation 
$\psi = e^{-iS}\tilde\psi$,
where $S$ was chosen to cancel the
interaction and is of the form
\beq
S(\{x_{+i}\}, \{x_{-i}\}) = \sum_{i,j}E(x_{+i}-x_{-j})
\quad {\rm where} \quad E(x) = \int_0^x dx' V(x').
\eeq
We can compute the difference between $S(x_{-i}=L)$ and $S(x_{-i} = 0)$
for any particular negative pseudospin coordinate $x_{-i}$ as 
\bea 
S(x_{-i}=L) -S(x_{-i} = 0) &=& \sum_{j} [E(x_{+j} -
L)-E(x_{+j})]\\
&=& n^T_+ \int_0^L V(x)dx \equiv n^T_+ \delta.
\eea
in terms of a phase shift $\delta$ and
$n^T_+$ which is the total number of positive pseudospin particles.  
One gets a similar result if we choose the reference
particle to be a positive pseudospin particle, with the only
difference that $n^T_+$ gets replaced by $n^T_-$ and $\delta$ by
$-\delta$.  Hence the quantisation condition on the wave-numbers of
the particles becomes 
\beq
Lk_{\pm i}\mp n^T_{\mp}\delta = 2\pi n_{\pm i}
\label{quant}
\eeq
where the $n_{\pm i}$ are integer quantum numbers analogous to those
used in the non-interacting case.
Since, in general, $n^T_{\mp}\delta
\ne$ integral multiple of $2\pi$, the free Hamiltonian and the 
interacting Hamiltonian are in different Hilbert spaces.

So far, all the arguments used by Schulz and Shastry have gone through
for our model as well. The differences begin when we try to construct
the many body ground state and the spectrum of low energy
excitations.  For ease of presentation, we will now 
specialise to the two band case, explicitly
perform the calculations leading to the low energy effective
Hamiltonian, and then generalise to the case of $N$ bands.

For the two-band case, the single particle dispersion is given by
\beq
{\tilde H} = e^{iS}He^{-iS} = \sum_{i \sigma}
{\large a_4 p_{\sigma i}^4 +a_2 p_{\sigma i}^2} 
\eeq 
where $a_4 = \Lambda^2$ has length dimension two and $a_2 = -1$ is
dimensionless. 
As mentioned before, we restrict the Fermi level to lie 
within the double well - $i.e.$, we have four distinct Fermi points  
($-F_2, -F_1$) on the left and ($F_1,F_2$) on the right.
For each energy,
the degeneracy is either four  or two , depending on whether or not both
$k_{1}$ and $k_{2}$ satisfies the boundary condition given in
Eq.(\ref{quant}). However, all that really 
matters is that energy levels in both the wells below the Fermi
level are filled. Let us assume that in band 1, there are $n_1$ states
below the Fermi level and in band 2, there are $n_2$ states below the
Fermi level for both pseudospins. The ground state energy is then given by
\beq 
E_{gs}= 2\sum_{\sigma=\pm}~~\sum_{i=-n_1}^{n_2}[ h_1 i^4 - h_2 i^2]
\eeq
where  $h_1 = \Lambda^2 ({2\pi\over L})^4$ and $h_2 = ({2\pi\over L})^2$. 
We have assumed that $n^T_{\pm}\dtpi$ is an integer and the factor
of two takes care of the contributions from both the right and left
moving sectors.
(Note that the states on the right branches of both the wells are
right movers, whereas the states on the left branches of both wells
are left movers. 
See figure below.)
\epsfxsize=5.8 in
\epsfysize=4.6 in
\vspace{.8 cm}

\begin{center}
\epsffile{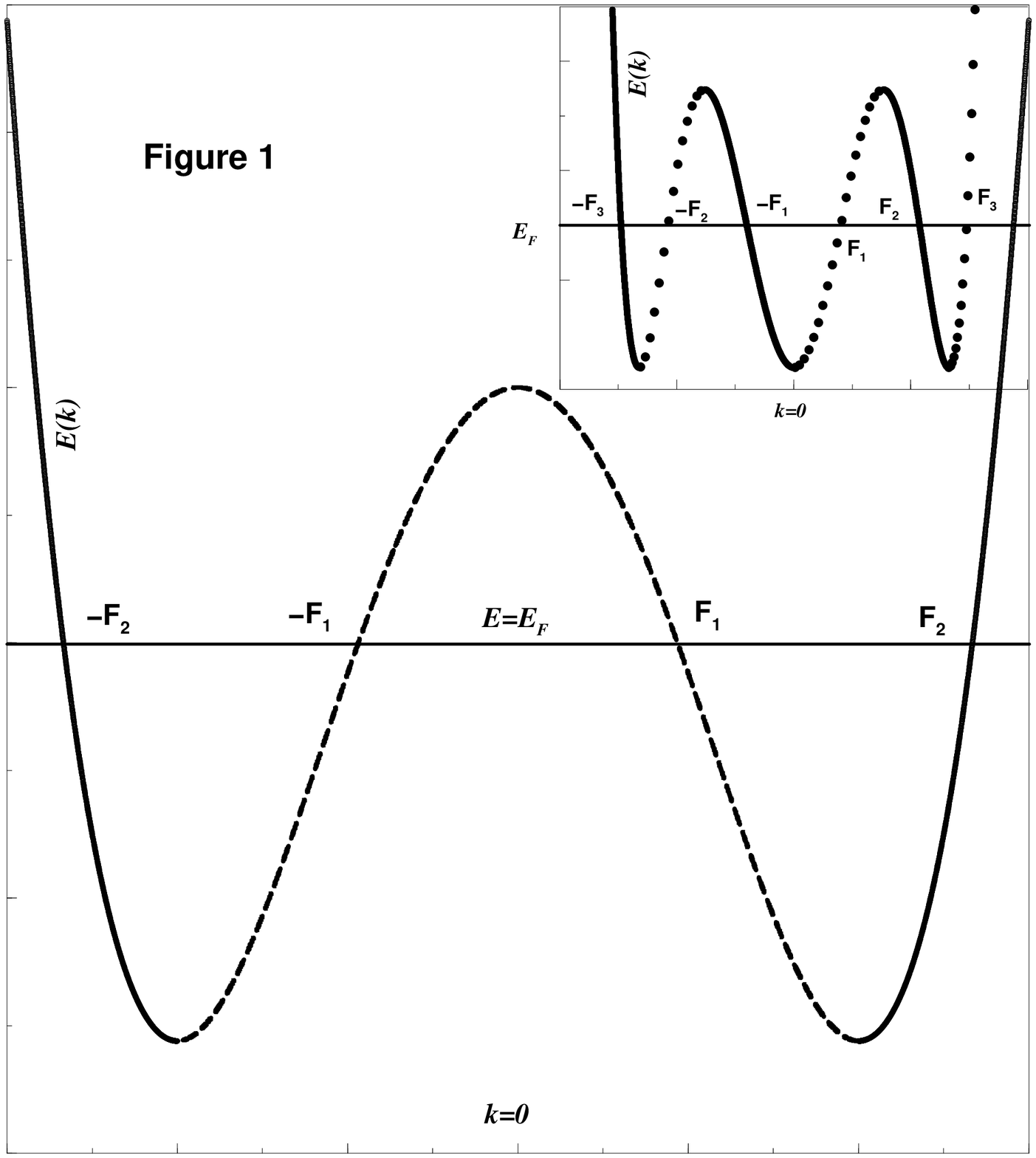}
\end{center}
\begin{itemize}
\item{ \bf Fig 1. } 
Dispersion for the two band model. For the Fermi energy
$E_F$, the two bands are denoted by dotted indices (band 1) and full
lines (band 2). $(-F_2, F_1)$ are clearly left-mover Fermi points
and $(-F_1, F_2)$ are right mover Fermi points. The inset shows the
dispersion of the three band model, with left-mover branches denoted
by full lines and right-mover branches denoted by dotted lines.
\end{itemize}

The second order fluctuation in the energy due to
the addition of $n_{R\pm I}$ and $n_{L\pm I}$ particles for the right
and left movers can also be computed. It is given by 
\bea
E^{(2)} &=& {\large [}\sum_{\pm}~~\sum_{-(n_1-n_{R\pm 1})}^{(n_2+n_{R\pm
2})}
[ h_1 (i+c_{\mp})^4 - h_2 (i+c_{\mp})^2] + R\longrightarrow L{\large ]}_2\\
&=& \sum_I^2\sum_{\pm}~~ [h_1f_1(n_I)+h_2f_2(n_I)]~~(n_{R\pm I} - c_{\mp})^2
\\
&=& \sum_{\pm}\sum_I^2 g(n_I)~~[J_{\pm I} \pm n_{\mp}\dtpi)^2 +
(n_{\pm I})^2]
\eea
where $c_{\sigma}=n^T_{\sigma}\dtpi$ in the first row and the
subscript 2 is to indicate that we keep only terms upto quadratic
order in the fluctuations. In the second row,
$f_1(n_I) = n_I(n_I+1)(2n_I+1)$
and $f_2=n_I+1/2$, and in the third row, we have
defined the current $J_{\pm I}=n_{R\pm I}-n_{L\pm I}$, the charge
$n_{\pm I} = n_{R\pm I} + n_{L\pm I}$ and the `density' $g(n_I) = 
h_1f_1(n_I)+h_2f_2(n_I)$.  The total charge 
is clearly  $n^T_{\pm}= n_{\pm 1} + n_{\pm 2}$. 
Note however, that unlike the Schulz-Shastry model, here fourth order
fluctuations do exist, which we neglect because we are only interested in
low energy fluctuations. 

We bosonise as in the single band case by
introducing boson fields $\phi_{\pm I}$ with their conjugate 
momenta $\Pi_{\pm I}$. These are related to the currents and charge 
densities as 
\bea
n_{\pm I} &=& {L\over \sqrt{\pi}}\partial_x \phi_{\pm I}\\
J_{\pm I} &=& -{L\over \sqrt{\pi}}\Pi_{\pm I}.
\eea
(We use the notation of Ref.\cite{SHANKBOS}.)
To rewrite the effective hamiltonian for the low energy fluctuations
in terms of the boson fields, we have to identify the function
$g(n_I)L/2\pi=\rho_I$ as an effective density after which we obtain
\beq
H = \sum_I^2\sum_{\pm}\int dx \rho_I\{[-\Pi_{I\pm} \pm {\delta\over\pi}
(\px\phi_{1\mp}+\px\phi_{2\mp}]^2 +(\px \phi_{\pm I})^2\}.
\label{htwo}
\eeq
But interestingly, although $\rho_I$
contains information about the scale, the low energy effective
Hamiltonian is scale invariant - there are no mass terms (or cosine
terms leading to mass terms) for the boson fields.
A similar redefinition of
variables as in the one band case, 
\beq
\tilde \phi_{\pm I} = \phi_{\pm I} ~~, ~~~~\tilde \Pi_{\pm I} =
\tilde\Pi_{\pm I} \mp {\delta\over\pi} (\px \phi_{\mp 1} +\px\phi_{\mp 2})
\eeq
leads to a non-interacting form of the
Hamiltonian given by
\beq
H = \sum_I^2\sum_{\pm}\int dx ~\rho_I~~[(\tilde\Pi_{\pm I})^2 + 
(\px \tilde\phi_{\pm I})^2 ~].
\label{htwon}
\eeq
Thus the correlators of the tilde fields are just free field correlators.
In terms of the non-tilde bosonic variables or equivalently
in terms of the fermion fields, the Hamiltonian is not
non-interacting. However, since they are explicitly known in terms of
the free fields, their correlators can also be explicitly calculated. 

In fact, at this stage, the generalisation to $N$ chains is obvious. 
The single particle dispersion of the $N$-band model has $N$ wells 
and $2N$ Fermi points. The Hamiltonian for quadratic fluctuations
about the Fermi points is precisely the same as that in Eq.(\ref{htwo})
with the replacement
\beq
(\px \phi_{\mp 1} +\px\phi_{\mp 2}) \rightarrow \sum_J^N\px \phi_{\mp J}
\eeq
As before, the
redefinition of $\phi_I$ and $\Pi_I$ in terms of the tilde fields
leads to the non-interacting form of the 
Hamiltonian in Eq.(\ref{htwon}) with the sum going over all $N$ bands.

We now compute correlation functions using the representation for the
fermion operators in terms of the non-interacting boson fields given
by
\bea
\psi_{R\pm I} &=& {\rm exp} (\tp_{R\pm I} \mp {\delta\over 2\pi}\sum_I^N
\tp_{\mp}), \\
\psi_{L\pm I} &=& {\rm exp} (\tp_{L\pm I} \pm  {\delta\over 2\pi}\sum_I^N
\tp_{\mp}).  
\eea
(We follow the notation in Ref.\cite{SHANKBOS} and define $\phi_{R\pm
I} = 1/2(\phi_{\pm I} - \int_{-\infty}^{\infty}\Pi_{\pm I}(x)dx)
$ and $\phi_{L\pm
I} = 1/2(\phi_{\pm I} + \int_{-\infty}^{\infty}\Pi_{\pm I}(x)dx)
$. ~)
The one-particle correlation function is given by
\beq
G_{Z\pm I}(x) = <\psi_{Z\pm I}(x)\psi_{Z\pm I}^{\dagger}(0)> \sim x^{-\eta} 
\eeq
with $\eta = 1+N(\delta^2/2\pi^2)$ for both right and left movers
($Z=R/L$), for both pseudospins and for all $I$.
As in the one-band case, the fermion has an anomalous dimension given
by $\eta\ne {\rm integer}$. 
This is the indication that the system is a Luttinger
liquid and not a Fermi liquid. The interesting point to note here is  the
dependence of the anomalous dimension on the number of chains. The
model is not just a collection of one-band  Luttinger liquids - there
exists a genuine dependence on the number of bands. 

We can also compute exponents of the two-particle operators(TPO) in order to
identify the relevant perturbations and hence potential
singularities. 
In the one-band case, the only non-trivial two-particle
correlations involved excitations at both the right and left Fermi
points, because these were the only two Fermi points. Here, however,
we can have non-trivial two-particle correlations involving excitations at
two right Fermi points and two left Fermi points as well. 
These exponents for the two particle correlations are tabulated below.
\vspace{.5 cm}
\begin {center}
{\bf \sc TWO PARTICLE CORRELATIONS}\\
\vspace{0.2cm}
\begin{tabular}{|l|l|l|l|} \hline 
TPO & $\eta $ &TPO($I\ne J$)& $\eta $ \\ \hline
$\psi_{R\pm I}^{\dagger} \psi_{L\pm J}$& 2&$\psi_{Z\pm I}^{\dagger} \psi_{Z\pm J}$& $2(1+N\dlts)$ \\ \hline
$\psi_{R\pm I}^{\dagger} \psi_{L\mp J}$& $2(1\mp \dlt)+N\dlts $&$\psi_{Z\pm I}^{\dagger} \psi_{Z\mp J}$&$2(1+\frac{N\delta^2}{2\pi^2})$ \\ \hline
$\psi_{R\pm I}\psi_{L\mp J}$& $2(1+N\dlts) $&$\psi_{Z\pm I}\psi_{Z\mp J}$& 2 \\ \hline
$\psi_{R\pm I}\psi_{L\mp J}$& $2(1\pm\dlt)+N\dlts $&$\psi_{Z\pm I}\psi_{Z\mp J}$&$2(1+\frac{N\delta^2}{2\pi^2})$  \\ \hline
\end{tabular}
\end{center}
\vspace{.5 cm}

Fortunately, they are independent of the band index and
only depend on whether they involve both right and left Fermi points
or right (left) movers at both Fermi points. Interestingly,
none of the RR or LL exponents lead to relevant perturbations. 
This is in agreement with the weak coupling RG approach\cite{SHANKAR},
where there is a non-zero contribution to the four-point vertex only
when there is momentum transfer between left and right Fermi points. 
In the single chain case considered in Ref.\cite{SHANKAR}, there was
no possibility of momentum transfers between two Fermi points on the
left or two on the right, since  the model only had one on each
side. However, even in the more general case of $N$ left-moving Fermi
points and $N$ right moving Fermi points\cite{MR}, graphs involving loop
momenta in two left-moving shells or two right moving shells are
always zero, because the energies have the same sign and the contour
integral for the energy vanishes. 

For positive $\delta$, the only relevant perturbations are
\bea 
\psi_{R+I}^{\dagger} \psi_{L-J} &\quad& \eta = 1+(1-\dpi)^2 +
(N-1) \ddpi, \nonumber \\
\psi_{R-I} \psi_{L+J} &\quad& \eta = 1+(1-\dpi)^2 +
(N-1) \ddpi.
\label{lastbo} 
\eea  
Clearly as the number of chains increases, the exponent increases
until at some critical value of $N=N_c \propto 1/\delta$, both the operators
above cease to be relevant. Similarly, for negative $\delta$, the
relevant perturbations are
\bea
\psi_{R-I}^{\dagger} \psi_{L+J} &\quad& \eta = 1+(1+\dpi)^2 +
(N-1) \ddpi, \nonumber \\
\psi_{R+I} \psi_{L-J} &\quad& \eta = 1+(1+\dpi)^2 +
(N-1) \ddpi.
\label{last}
\eea  
which cease to be relevant beyond $N_c$. Hence, for $N>N_c$
chains, there are no relevant perturbations at all. The system is
always a Luttinger liquid. For $N<N_c$, the system, depending on 
which instability grows, (which perturbation will be added),  which in turn, 
will be dictated by the realistic model that we wish to study, 
will be in different ground states.

Let us compare our results with the results obtained by giving
additional internal degrees of freedom to the $\sigma = \pm$
particles\cite{SS}. If we assume that they occur in $m$-flavours, then 
the Hamiltonian is just
\beq
H =  \sum_{I=1}^m \sumsig a_{I} (\pisigm)^{2}.
\eeq
Surprisingly, an analogous calculation leads precisely to 
the same exponents as in
Eqs.(\ref{lastbo}) and (\ref{last}) with $N$ replaced by $m$. 
However, in this case, there genuinely exist $2m$ degrees of freedom,
and the various two particle correlators have physical meaning. $I=J$
give two particle correlators of the same particle, whereas there is
no analogue of $I\ne J$ correlators. The exponents are actually
independent of the particle index because of the internal symmetry.
For our Hamiltonian in Eq.(1), however, there are only two 
degrees of freedom correponding to the $\sigma = \pm$ particles. 
It is only after linearising around the different Fermi points and assuming
that each of the linearised fermions can be bosonised independently
that we have $N$ independent R/L moving fermions or bosons, whose
correlators can be computed independently.  For the original fermions,
the only relevant charges are  $n^T_{\pm}$ and the relevant currents
are $J_{\pm} = \sum_I (n_{R\pm I} - n_{L\pm I})$. 

Many of the issues in coupled chain models, however, remain
unaddressed in this rather simple model, which is perhaps better
thought of as a single chain model with a more complicated band
structure. To really apply this model to $N$-chains, one would have to
modify the model, so that there is some analog of the interplay between
interchain hopping and intra-chain interactions.
However, note that even as an $N$-band model, 
it is not trivial that
the correlation  functions are identical to those  of the $m$- flavour
model. 

In conclusion, we have studied a general model with $2N$ Fermi points,
(an $N$-band or $N$ chain model), which is exactly solvable and has
non-trivial Luttinger liquid behaviour. We computed the exponents of
the various two-particle operators and found the possible relevant
perturbations. Interestingly, we found that the exponents have
non-trivial dependence on the number of chains - they are not
merely additive. Furthermore, beyond a certain number of chains $N_c$, which is
inversely proportional to the strength of the interaction $\delta$,
all perturbations are irrelevant and the Luttinger liquid ground state
is robust. Thus, this model, if it can be effectively generalised to 
higher dimensions would be a good
starting point to study possible Luttinger liquid ground states in
higher dimensions.

\section*{Acknowledgments}
RKG would like to thank  the Mehta Research Institute
for hospitality during the course of this work.

\end{document}